\documentclass[superscriptaddress,aps,twocolumn]{revtex4-2}

\usepackage{graphicx}
\usepackage{dcolumn}
\usepackage{bm}
\usepackage{amsmath,amssymb,amsfonts}%
\usepackage{amsthm}%
\usepackage{mathrsfs}%
\usepackage[colorlinks=true,citecolor=blue,linkcolor=blue]{hyperref}

\usepackage{float}
\begin{document}

\preprint{APS/123-QED}

\title{Chirped-Pulse Forward Raman Amplification in Nonuniform Plasmas}

\author{Zhi-Yu Lei}
\affiliation{National Key Laboratory of Dark Matter Physics, School of Physics and Astronomy, Shanghai Jiao Tong University, Shanghai 200240, China}
\affiliation{Laboratory for Laser Plasmas and Collaborative Innovation Centre of IFSA, Shanghai Jiao Tong University, Shanghai 200240, China}

\author{Zheng-Ming Sheng}
\email{Corresponding author: zmsheng@sjtu.edu.cn}
\affiliation{National Key Laboratory of Dark Matter Physics, School of Physics and Astronomy, Shanghai Jiao Tong University, Shanghai 200240, China}
\affiliation{Laboratory for Laser Plasmas and Collaborative Innovation Centre of IFSA, Shanghai Jiao Tong University, Shanghai 200240, China}
\affiliation{Tsung-Dao Lee Institute, Shanghai Jiao Tong University, Shanghai 201210, China}

\author{Su-Ming Weng}
\email{Corresponding author: wengsuming@sjtu.edu.cn}
\affiliation{National Key Laboratory of Dark Matter Physics, School of Physics and Astronomy, Shanghai Jiao Tong University, Shanghai 200240, China}
\affiliation{Laboratory for Laser Plasmas and Collaborative Innovation Centre of IFSA, Shanghai Jiao Tong University, Shanghai 200240, China}

\author{Min Chen}
\affiliation{National Key Laboratory of Dark Matter Physics, School of Physics and Astronomy, Shanghai Jiao Tong University, Shanghai 200240, China}
\affiliation{Laboratory for Laser Plasmas and Collaborative Innovation Centre of IFSA, Shanghai Jiao Tong University, Shanghai 200240, China}

\author{Jie Zhang}
\affiliation{National Key Laboratory of Dark Matter Physics, School of Physics and Astronomy, Shanghai Jiao Tong University, Shanghai 200240, China}
\affiliation{Laboratory for Laser Plasmas and Collaborative Innovation Centre of IFSA, Shanghai Jiao Tong University, Shanghai 200240, China}
\affiliation{Tsung-Dao Lee Institute, Shanghai Jiao Tong University, Shanghai 201210, China}

\date{\today}

\begin{abstract}

Light amplification via Raman scattering in plasma has been severely constrained by stringent phase matching conditions and the need for plasma uniformity. To overcome these limitations, we propose a forward Raman amplification scheme that employs a positively chirped seed pulse co-propagating with a pump pulse in a nonuniform plasma with an upramp density profile. We demonstrate that the phase detuning induced by plasma nonuniformity can be dynamically compensated, enabling broadband amplification across the entire spectral bandwidth of the seed pulse. Concurrently, the chirped pulse duration undergoes continuous compression as a result of the spatially varying dispersion of the plasma. Our theoretical model, incorporating the detuning term and supported by particle-in-cell simulations, elucidates the compensation mechanism. It is shown that a chirped seed pulse with an initial bandwidth $\sim10\%$ can be directly amplified by a factor of $10^7$ to an intensity exceeding $10^{17}~\text{W/cm}^2$ within a picosecond timescale in a steep density ramp. This scheme establishes a new foundation for advancing plasma-based light amplification toward practical applications.

\end{abstract}

\maketitle

Plasma optics has emerged as a promising platform for manipulating extremely high-intensity laser pulses, owing to its ability to sustain energy densities far beyond the damage thresholds of solid-state optical materials~\cite{stuart1995laser,wu2005chirped,wu2005manipulating,edwards2021laser,riconda2023plasma}. Within this framework, the three-wave coupling mechanism offers a viable route to light amplification in plasma, including backward Raman amplification (BRA)~\cite{shvets1998superradiant,malkin1999fast,trines2011simulations,lehmann2013pulse,vieira2016amplification}, forward Raman amplification (FRA)~\cite{lei2025towards,wu2020compression}, and the strong-coupling backward Brillouin amplification (sc-BBA)~\cite{andreev2006short,lehmann2013nonlinear,weber2013amplification,chiaramello2016role,wu2024efficient}. The plasma waves generated in these processes---specifically, electron plasma waves (EPWs) in stimulated Raman scattering and ion-acoustic waves (IAWs) in stimulated Brillouin scattering---act as a continuous mediator for energy transfer from the pump pulse to the seed pulse.

In Raman-based schemes, the high-frequency EPWs offer the advantages of high growth rates and the capability to amplify long-wavelength pulses. Nevertheless, several challenges remain, including low thresholds for detrimental kinetic effects in backward configurations~\cite{hur2005electron,edwards2015efficiency} and stringent requirements for plasma uniformity due to high sensitivity to phase-matching fluctuations. These factors have restricted the amplification efficiency in BRA experiments to below $10\%$~\cite{ping2004amplification,ren2007new,vieux2017ultra,wu2019stimulated}, with the maximum reported output power still below $0.1$ terawatt~\cite{edwards2021laser}. To address these issues, a number of theoretical and experimental studies have employed chirped pump pulses to tailor the gain distribution and mitigate kinetic saturation~\cite{ersfeld2005superradiant,vieux2011chirped,yang2015chirped}; however, the amplification has remained in the linear regime, yielding only modest gain. Chirped seed pulses have also been adopted in BRA at relatively high plasma densities, enabling maximum amplification over a reduced propagation length.~\cite{toroker2012seed}. All of the aforementioned BRA schemes assume a uniform plasma distribution to satisfy ideal phase-matching conditions. In contrast, Brillouin schemes benefit from the low-frequency IAWs, which render them immune to electron kinetic effects and plasma nonuniformity~\cite{edwards2016short,chiaramello2016optimization}, thereby achieving an energy transfer efficiency of $20\%$ in experiments~\cite{marques2019joule}. On the other hand, the same low-frequency IAWs also limit the growth rate of sc-BBA, reducing both the amplification gain and the achievable pulse compression. Moreover, Brillouin amplification is restricted to seed pulses whose frequencies are nearly identical to that of the pump pulse.

In this Letter, we propose a distinct mechanism for plasma-based light amplification based on forward Raman amplification with a chirped seed pulse (cFRA) in nonuniform plasma. Unlike conventional approaches that strive to maintain perfect phase matching throughout the interaction, the cFRA scheme intentionally operates under detuned conditions. By replacing a narrowband seed pulse with a chirped one that possesses a substantially broader frequency spectrum, the scheme enables successive compensation for phase mismatches caused by plasma-frequency variations arising from plasma inhomogeneity and the fixed central frequency of the pump pulse. The cFRA scheme retains the key advantages of co-propagating FRA~\cite{lei2025towards}---namely robustness against kinetic effects, high amplification rates, and experimental feasibility---while simultaneously achieving higher efficiency and simpler implementation than standard FRA.

\begin{figure}[t]
\centering
\includegraphics[width=0.48\textwidth]{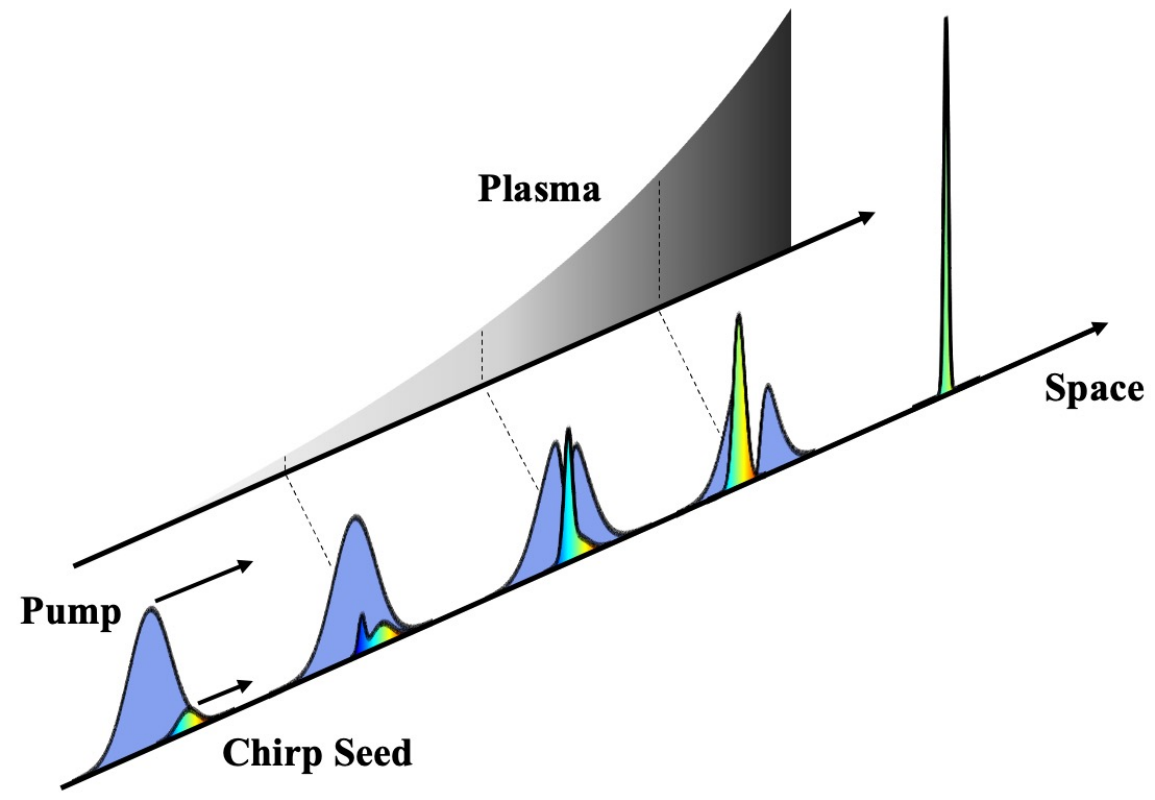}
\caption{Concept of the cFRA scheme, where the pump pulse, the positively chirped seed pulse, and the plasma density gradient are along the same direction. In the low density region, the high frequency components of the seed initially located at its trailing edge are amplified first. As the pump gradually overtakes the seed in the high density region, the low frequency components of the seed located the front of the seed begin to be amplified successively. During the propagation and amplification process, the duration of the amplified seed pulse is compressed simultaneously.}
\label{model}
\end{figure}

Figure~\ref{model} shows the basic concept of the cFRA scheme. In the model, both a positively chirped seed pulse and the pump pulse co-propagate along the same direction in the inhomogeneous plasma with increasing density. Initially, when the two pulses interact in the low-density region with a low-plasma frequency, the trailing edge of the seed pulse with high frequency components is amplified first,  where the corresponding phase matching conditions are satisfied dynamically. As the pump, which has a higher frequency and thus a higher group velocity than the seed, continuously overtakes the seed pulse and propagates towards higher density regions, low frequency components of the chirped seed located at its leading edge start to be amplified. Meanwhile, as the trailing  edge of the chirped seed with high frequency components propagates with an overall higher group velocity than its leading edge, the seed rear is chasing its front, causing seed pulse compression. Eventually, in the presence of the density variation, the cFRA scheme effectively amplifies all frequency components of the chirped seed pulse and compresses them in duration. Ultimately, cFRA enables the seed pulse to achieve high peak intensity and extreme short output duration, while also allowing rescaling to shorter interaction time and plasma dimensions without requiring strict density uniformity.

We start by presenting an analytical model of cFRA. Assume that both the pump and chirped seed lasers co-propagate in an inhomogeneous plasma along the $+x$ direction, the corresponding coupled three-wave equations with detuning included are then given by
\begin{align}
    (\partial_t+v_0(x)\partial_x)a_0&=\frac{ck_2(x)}{4}\sqrt{\frac{\omega_{\mathrm{pr}}}{\omega_0}}a_1a_2\label{3w-1},\\
    (\partial_t+ v_1(x)\partial_x)a_1&=-\frac{ck_2(x)}{4}\sqrt{\frac{\omega_{\mathrm{pr}}\omega_0}{\omega_1^2}}a_0a_2^*\label{3w-2},\\
    (\partial_t+i\delta\omega(x,t))a_2&=-\frac{ck_2(x)}{4}\frac{\omega_{\mathrm{pr}}}{\omega_2(x)}\sqrt{\frac{\omega_{\mathrm{pr}}}{\omega_0}}a_0a_1^*\label{3w-3},
\end{align}
where $\delta\omega=\omega_2-\omega_0+(\omega_1+d\phi_{\mathrm{ch}}/dt)$ is the detuning from the three-wave resonance. Here the index $0$, $1$, $2$ represents the pump, seed, and electron plasma waves, respectively, $a_{0,1,2}$ are the normalized envelope amplitudes, $v_{0,1}$ are the group velocities, $k_2$ is the wave vector of electric plasma waves, $\omega_{0,1,2}$ are the frequencies, $\omega_{\mathrm{pr}}=\omega_0-\omega_1$ is the resonant plasma frequency, and $\phi_{\mathrm{ch}}$ is the extra phase introduced by the chirped seed pulse. In above equations, we assume a cold plasma case where there is no dispersion for electron plasma waves, leading to their wave vector always satisfying the resonance condition as the density changes by $k_2=k_0(x)-k_1(x)$~\cite{malkin2000detuned}. Thereby, only net detuning on the frequency term is included while the wave vectors are perfectly matched during the amplification. To solve the equations analytically, we treat the chirped seed as the combination of an infinite number of independent frequency components $\omega_{1,n}$, where $n$ is the serial number for the components. For the amplification of seed component $\omega_{1,n}$, the corresponding wave vector of the excited electron plasma wave is $k_{2,n}$, and the frequency detuning is $\delta\omega_n=\omega_2(n_e)-\omega_0+\omega_{1,n}$ related to the plasma density distributions.

We then consider the linear stage where the pump depletion is negligible with a constant value. By ignoring the $\partial_x$ term compared with the $\partial_t$ term for simplicity, the coupling equations for the $n$th component become $a_0=a_{0,0},\,\partial_t a_{1,n}=c_1a_2^*,\,\,(\partial_t+i\delta\omega_n)a_2=c_2a_{1,n}^{*}$, where $c_1=-a_{0,0}ck_{2,n}\sqrt{\omega_{\mathrm{pr},n}\omega_0}/4\omega_{1,n}$ and $c_2=-a_{0,0}ck_{2,n}\omega_{\mathrm{pr},n}\sqrt{\omega_{\mathrm{pr},n}/\omega_0}/4\omega_2$, with $\omega_{\mathrm{pr},n}=\omega_0-\omega_{1,n}$. Based upon above simplified coupling equations, we define $g$ as the effective growth rate given by
\begin{equation}
    g(x)=\sqrt{c_1(x)c_2(x)-\delta\omega_n^2(x)/4}.
    \label{geff}
\end{equation}
The amplitude of the seed pulse with frequency component $\omega_{1,n}$ in the linear amplification stage can then be solved analytically as
\begin{equation}
    a_{1,n}=\left\{
    \begin{aligned}
            &a_{1,0}\mathrm{cosh}(gt) \mathrm{exp}(i\phi_{1,0}),\,\,\,\delta\omega_n=0\\
            &\frac{a_{1,0}}{\sqrt{2}}\left(\alpha_n\mathrm{cosh}(2gt)+\beta_n\right)^{1/2} \mathrm{exp}(i\Phi_1),\,|\delta\omega_n|<2g_{\mathrm{FRA}}  \\
            &\frac{a_{1,0}}{\sqrt{2}}\left(\alpha_n\mathrm{cos}(2|g|t)+\beta_n\right)^{1/2} \mathrm{exp}(i\Phi_1),\,|\delta\omega_n|>2g_{\mathrm{FRA}}
            \label{a1}
    \end{aligned}
    \right.
\end{equation}
Here, $\alpha_n=1+(\delta\omega_n/2g)^2$, $\beta_n=1-(\delta\omega_n/2g)^2$, and $\Phi_1=\phi_{1,0}+\delta\omega_nt/2-\mathrm{tan}^{-1}[\delta\omega_n\mathrm{tanh}(gt)/2g]$, $g_{\mathrm{FRA}}=\sqrt{c_1c_2}=a_{0,0}ck_2\sqrt{\omega_{\mathrm{pr}}/(\omega_0-\omega_{\mathrm{pr}})}/4$ corresponding to the linear growth rate for perfect phase matching with $\delta \omega_n=0$~\cite{lei2025towards}.  Note that $g$ is a imaginary number for $|\delta\omega_n|>2g_{\mathrm{FRA}}$ and correspondingly $\Phi_1=\phi_{1,0}+\delta\omega_n t/2-\mathrm{tan}^{-1}[\delta\omega_n \mathrm{tan}(|g|t)])/2|g|]$. When the phase matching conditions are not satisfied, Eq.~\eqref{a1} shows that the seed pulse can still be amplified effectively by $\mathrm{cosh}(2gt)$ as long as the detuning $\delta\omega_n$ is controlled between $-2g_{\mathrm{FRA}}$ to $2g_{\mathrm{FRA}}$; The larger the typical linear growth rate of FRA, the greater the detuning the amplification system can tolerate. On the other hand, if the detuning becomes too large, the amplitude of the seed pulse then shall oscillate as a cosine function without sustained amplification. Meanwhile, based on the global phase model which is primarily discussed in BBA schemes~\cite{chiaramello2016role}, the extra phase terms of the amplified seed pulse introduced by the detuning also suppress the energy reversal, leading to higher efficiency in the cFRA scheme.


\begin{figure}[t]
\centering
\includegraphics[width=0.43\textwidth]{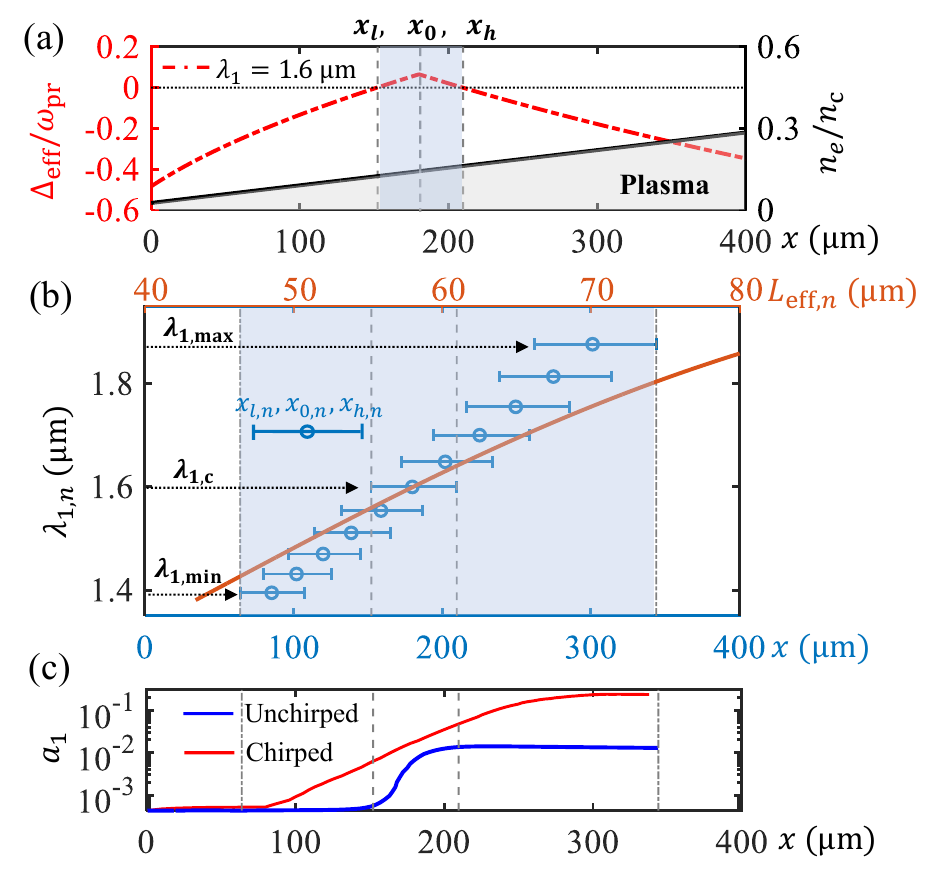}
\caption{Theoretical results of cFRA scheme. (a) The effective amplification region of a unchirped seed pulse in the nonuniform plasma, where the red dotted-dash line is related to the effective growth rate, and the black line is the plasma density distribution. (b) The effective amplification region (blue bars with circles) and the amplification length (red line) of each wavelength components of a chirped seed pulse in nonuniform plasma. (c) The evolution of the seed intensity solved by 3-wave coupling equations numerically.}
\label{theory}
\end{figure}
Based upon above analysis, we consider a typical case for the plasma density profile along the propagation distance with $n_e=n_0(1+(x-x_0)/L)$, where $n_0=0.1425n_{c0}$ is the resonant density, $n_{c0}$ is the critical density for the pump pulse, $x_0=180\,\mathrm{\mu m}$ is the resonant point, and $L=220\,\mathrm{\mu m}$ is the density scale length. With the nonuniform plasma, we first consider a case with a unchirped seed pulse having central wavelength $\lambda_1=1.6\,\mathrm{\mu m}$, initial intensity $I_1=10^{11}\,\mathrm{W/cm^2}$, duration $\tau_1=40T_0$, pumped by a pulse with wavelength $1.0\,\mathrm{\mu m}$, intensity $I_0=3\times 10^{16}\,\mathrm{W/cm^2}$, and duration $\tau_0=80T_0$. Figure \ref{theory}(a) shows that only a narrow amplification window is found from $x_l$ to $x_h$ around the resonant point $x_0$ (the blue area in Fig.~\ref{theory}(a)) in this case, where the effective growth rate is real and positive, defined as $\Delta_{\mathrm{eff}}=c_1c_2-\delta\omega_n^2/4>0$. However, if the seed pulse is linearly chirped by $\omega_{1,n}=w_1+\beta(t_n-\tau_1)$, $n=1,2,3...$, where $\beta=1.3\times 10^{27}\,\mathrm{rad/s^{2}}$ is the linear chirp parameter, then each independent component $\lambda_{1,n}$ from $\lambda_{1,\mathrm{min}}=1.39\,\mathrm{\mu m}$ to $\lambda_{1,\mathrm{max}}=1.87\,\mathrm{\mu m}$ satisfies distinct effective amplification conditions given by $|\delta\omega_n|<2g_{\mathrm{FRA},n}$, where $g_{\mathrm{FRA},n}=a_{0,0}ck_{2,n}\omega_{\mathrm{pr},n}/4\sqrt{\omega_{1,n}\omega_2}$, $\delta\omega_n=\omega_2-\omega_0+\omega_{1,n}$, $k_{2,n}=(\sqrt{\omega_0^2-\omega_2^2}-\sqrt{\omega_{1,n}^{2}-\omega_2^2})/c$, $\omega_{\mathrm{pr},n}=\omega_0-\omega_{1,n}$, and $\omega_2=\sqrt{n_ee^2/\epsilon_0 m_e}$, leading to each amplification window ranging from $x_{l,n}\approx  [(-2g_{\mathrm{FRA},n}/\omega_{\mathrm{pr},n}+1)^2-1]L+x_{0,n}$ to $x_{h,n}\approx[(2g_{\mathrm{FRA},n}/\omega_{\mathrm{pr},n}+1)^2-1]L+x_{0,n}$ at its resonant point $x_{0,n}=\{[\omega_{\mathrm{pr},n}/(\omega_0-\omega_1)]^2-1\}L+x_0$, as shown in Fig.~\ref{theory}(b). When the wavelength of the seed component increases, the corresponding resonant point and the length of the amplification window $L_{\mathrm{eff},n}=x_{h,n}-x_{l,n}=8g_{\mathrm{FRA},n}L/\omega_{\mathrm{pr},n}$ both increase. Thereby, the total effective amplification region is enlarged due to the contribution of each independent amplification window for different seed components. This indicates that the chirped seed pulse will be amplified by different frequency components successively as it enters different plasma density regions. To further illustrate above theoretical model, we have numerically solved the coupling equations and tracked the seed pulse amplitude as it propagates in the plasma. As shown in Fig.~\ref{theory}(c), the numerical results of the growth regions for both chirped and unchirped cases are consistent with the theoretical predictions, and the chirped seed pulse can obtain more sufficient amplification in the nonuniform plasma via cFRA. 

\begin{figure*}[t]
\centering
\includegraphics[width=1.0\textwidth]{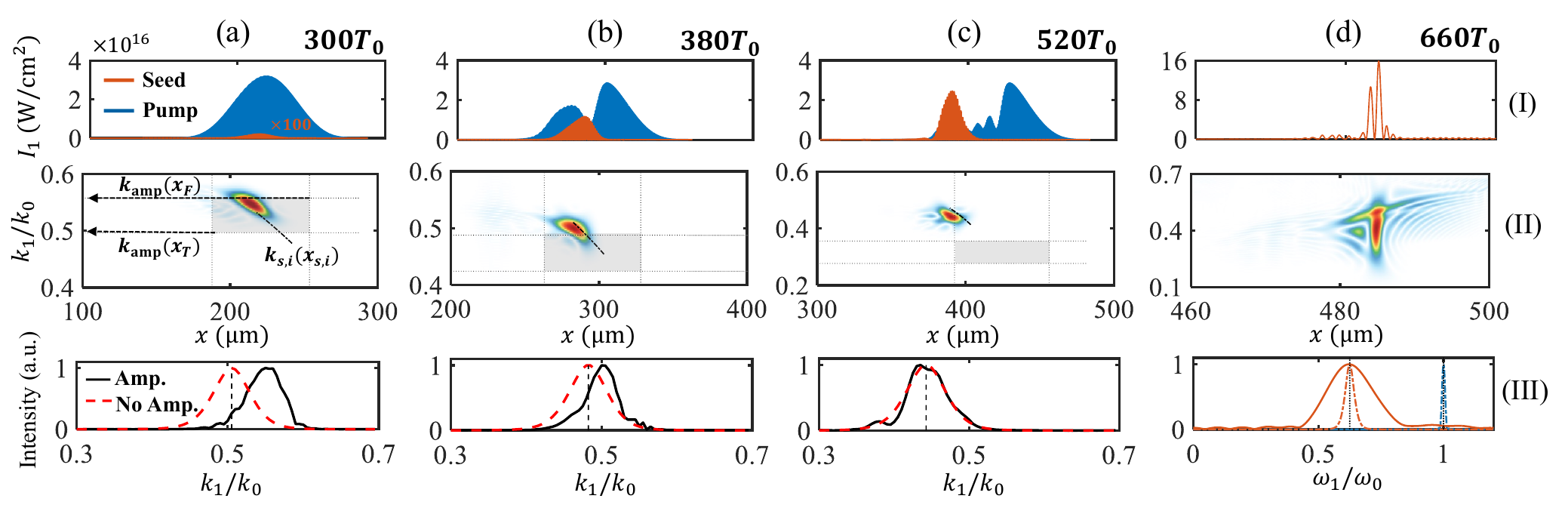}
\caption{The snapshots of the chirped seed pulse during amplification given in columns (a), (b), and (c). The first row (I) represents the intensity distributions of seed (red) and pump (blue) pulses. The second row (II) is the wave vector distributions of the seed pulse, where the grey areas mark the amplification regions determined theoretically according to the locations of the pump front ($x_F$) and tail ($x_T$) at the given time, and the corresponding plasma densities. The black dashed lines mark the theoretical wave vector trajectory of the chirped seed. The final row (III) is the wave vector spectra of the chirped seed pulse (black solid line) in plasma during amplification, where the red dashed lines are for the corresponding chirped seed pulse without amplification (i.e., without the pump).  Column (d) shows the frequency spectra of the incident two pulses and the output seed pulse after propagating through plasma in vacuum.}
\label{1d}
\end{figure*}
To further validate our theory, particle-in-cell (PIC) simulations have been carried out. Figure~\ref{1d} presents the 1D PIC results obtained by EPOCH~\cite{arber2015contemporary} to demonstrate the cFRA scheme. The pump pulse has wavelength $1.0\,\mathrm{\mu m}$, peak intensity $3\times10^{16}\,\mathrm{W/cm^2}$, and duration $233\,\mathrm{fs}$. The chirped seed pulse initially has the central wavelength $1.6\,\mathrm{\mu m}$, duration $116\,\mathrm{fs}$, chirp rate $\beta=0.6\times10^{27}\,\mathrm{rad/s}^{2}$ (corresponding to a total bandwidth $\Delta\omega/\omega\approx5.91\%$, which theoretically can be obtained from a $37\,\mathrm{fs}$ laser pulse and accessible in the experiments~\cite{sekhar202320,xia2026generation}), and peak intensity $10^{12}\,\mathrm{W/cm^2}$. The parameters of the background plasma with a linear upramp density profile are given by $n_0=1.56\times 10^{20}\,\mathrm{cm}^{-3}$, $x_0=250\,\mathrm{\mu m}$, and $L=400\,\mathrm{\mu m}$. As shown in column (a) in Fig.~\ref{1d} for the early stage of cFRA, comparing the spectrum of the chirped seed pulse during amplification with that of the seed pulse without pump amplification reveals that the amplification of the chirped seed mainly occurs in the high frequency components at this stage, resulting in an asymmetrically sharp peak in the spectrum. As the pulses propagate to the higher density regions, more lower frequency components are amplified, and the peak of the spectrum starts to shift towards the center, as can be found in columns (b) and (c) in Fig.~\ref{1d}. Unlike most light amplification schemes, in which the pulse may experience pulse duration broadening due to limited gain bandwidth, the cFRA scheme maintains broad-band amplification through a dynamic phase detuning compensation mechanism. Meanwhile, by tracking the theoretical wave vector trajectory of each frequency component $k_{1,n}(x_{1,n})$, where $k_{1,n}=\omega_{1,n}/c\sqrt{1-n_e(x_{1,n})/n_{c1,n}}$ and $x_{1,n}=c\sqrt{1-n_e(0)/n_{c1,n}}(t-t_n)-n_0c^2(t-t_n)^2/4n_{c1,n}L$ are the wave vector and the corresponding location of the $n\,\mathrm{th}$ component, one finds that the amplification is triggered sequentially from the tail to the front of the seed pulse during its propagation, consistent with the theoretical prediction, and that the duration of the seed pulse is compressed continuously due to plasma dispersion. Once the seed pulse exits both of the amplification region and the pump interaction region, it undergoes further self-compression into nearly single cycle due to nonlinear plasma effects, with its intensity boosted beyond $10^{17}\,\mathrm{W/cm^2}$ while its central frequency remains unchanged, as shown in column (d) of Fig.~\ref{1d}. 

\begin{figure}[b]
\centering
\includegraphics[width=0.37\textwidth]{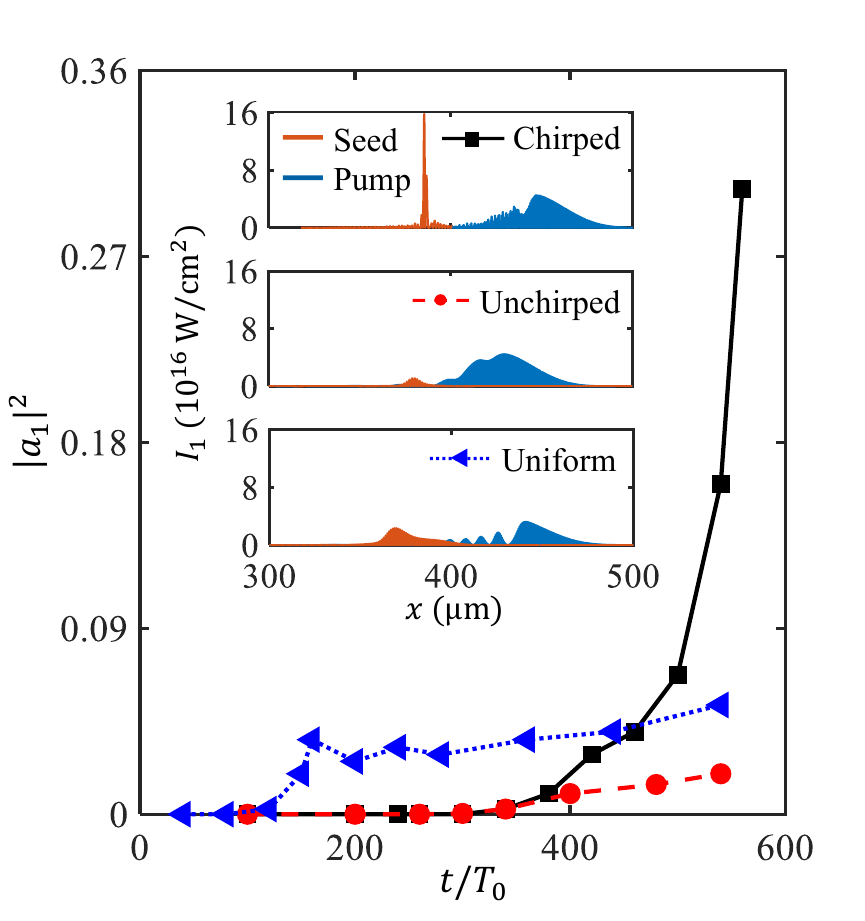}
\caption{The peak intensity of the seed pulse as a function of time found from PIC simulations for three cases, i.e., the chirped seed in nonuniform plasma (black line), unchirped seed in nonuniform plasma (red line), and unchirped seed in uniform plasma (blue line). The inset plots are the output intensity distributions at the same time for the three cases.}
\label{3cases}
\end{figure}
We then compare three different cases of seed pulse amplification, i.e., the chirped seed pulse in nonuniform plasma, unchirped seed in nonuniform plasma, and unchirped seed in uniform plasma. As shown in Fig.~\ref{3cases}, the seed pulse in the chirped case is amplified seven orders of magnitude from $10^{10}\,\mathrm{W/cm^2}$ to $2\times 10^{17}\,\mathrm{W/cm^2}$, and its duration is compressed to nearly single cycle within $10\,\mathrm{fs}$. The whole interaction time is within $1.8\,\mathrm{ps}$ and the required plasma length is less than $400\,\mathrm{\mu m}$. Meanwhile, the cFRA scheme can sustain much longer amplification time compared with the unchirped one, as the intensity of the unchirped seed pulse is found to saturate quickly at $10^{15}\,\mathrm{W/cm^2}$ in the same inhomogeneous plasma. Because the chirped seed can successively compensate for the mismatching between laser central frequency and the varying plasma frequency, the cFRA scheme can effectively mitigate the amplification truncation caused by the phase detuning. Although the amplification in cFRA is delayed compared with the typical FRA scheme in uniform plasma, where a perfect phase matching conditions are satisfied, both cases show a similar growth rate. As a unique feature, the energy reversal from the seed to the pump is largely suppressed in cFRA due to the varying Langmuir wave frequency and chirp-induce dispersion effects in the inhomogeneous plasma. Consequently, in the cFRA scheme, the seed pulse can undergo continuous amplification rather than oscillating in the nonlinear stage after $180T_1$ as observed in conventional FRA. This enables a shorter plasma length and interaction time, mitigating the growth of potentially negative kinetic effects while maintaining high efficiency and compactness. 

To further show the robustness of the cFRA scheme, some two-dimensional (2D) and three-dimensional (3D) PIC simulation results are given in the End Matter of this letter, where a PW-class output via cFRA is demonstrated. Kinetic effects such as wave-breaking, Landau damping, filamentation, and self-focusing are not prominently observed, and the energy loss due to ionization and collisional heating remains limited in the forward Raman scheme~\cite{lei2025towards}, indicating the strong robustness of cFRA. Meanwhile, the backward scattering of the pump by thermal fluctuations is naturally stabilized by the plasma density gradient~\cite{clark2003operating}.

\begin{figure}[t]
\centering
\includegraphics[width=0.48\textwidth]{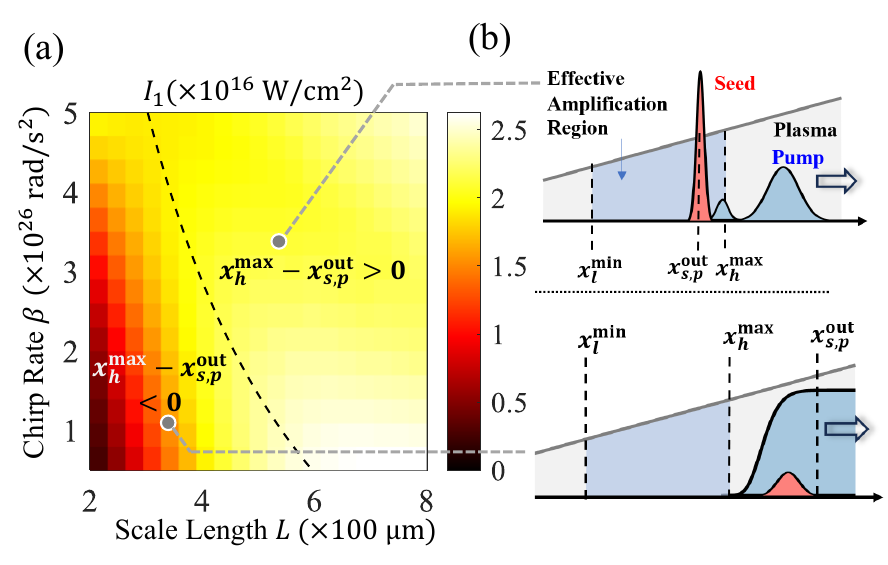}
\caption{(a)The parameter window of the density scale length $L$ and the chirp rate $\beta$ suitable for the cFRA scheme. The dashed line is the theoretical prediction given by $x_h^{\mathrm{max}}-x_{s,p}^{\mathrm{out}}=0$, where its right side corresponds to the full amplification of cFRA. The chirped seed pulse has a central wavelength of $1.7\,\mathrm{\mu m}$, an intensity of $10^{11}\,\mathrm{W/cm^2}$, and a duration of $40T_0$, while the parameters for the pump pulse is $1.0\,\mathrm{\mu m}$, $3\times 10^{16}\,\mathrm{W/cm^2}$, and $80T_0$. (b) Schematic plots of two light amplification cases indicated in panel (a). }
\label{parameters}
\end{figure}
To guide the future experimental studies of the cFRA scheme, we present a suitable parameter window for the density scale length $L$ and the chirp rate $\beta$. One criterion for achieving sufficient amplification via cFRA is that the detachment position between two pulses $x_{s,p}(t_{\mathrm{out}})$ must lie within the upper boundary of the total effective amplification region $x_{h}^{\mathrm{max}}$. Otherwise, the pump depletion becomes insufficient, causing the seed pulse to remain in interaction with the pump even after traversing the entire effective amplification region. Based on above considerations, we have numerically solved a series of three-wave coupling equations with detuning included under different parameters. Meanwhile, a critical line demarcating the sufficient and insufficient amplification regions is theoretically obtained from $x_h^{\mathrm{max}}-x_{s,p}(t_{\mathrm{out}})=f(\beta, L)=0$, as shown by the dashed curve in Fig.~\ref{parameters}. Here $t_{\mathrm{out}}$ is the solution to $x_s-x_p=0$, where $x_{s,p}(t)=c\sqrt{1-n_e(0)/n_{c}^{s,p}}(t-2\tau_{s,p})-n_0c^2(t-2\tau_{s,p})^2/4n_{c}^{s,p}L$, and $x_{h}^{\mathrm{max}}\approx[(2g_{\mathrm{FRA}}^{\mathrm{max}}/\omega_{\mathrm{pr}}^{\mathrm{max}}+1)^2-1]L+x_0^{\mathrm{max}}$ related to the chirp rate. As shown in Fig.~\ref{parameters}, on the left side of the critical line (i.e., $x_{s,p}(t_{\mathrm{out}})>x_{h}^{\mathrm{max}}$), the amplified intensity under the corresponding parameters is significantly lower than that on the right side (i.e., $x_{s,p}(t_{\mathrm{out}})<x_{h}^{\mathrm{max}}$). Nevertheless, as the parameter values approach the critical line, the intensity grows continuously and tends to saturate after crossing the line, indicating sufficient amplification. Additionally, when the plasma density distribution becomes steeper (i.e., smaller $L$), the required chirp rate for the sufficient amplification increases accordingly. This is because a larger phase detuning caused by the density change requires a broader range of frequency components to compensate. These results clearly demonstrate the effectiveness of our theoretical model, thereby providing a concise reference for the experimental design of cFRA, and guiding the selection of optimal laser parameters based on accessible plasma conditions in experiments.

In conclusion, we have proposed the cFRA scheme, which effectively converts the detrimental effect of plasma nonuniformity on Raman amplification into a beneficial one. By introducing a chirped seed pulse with a broader spectrum of frequency components, one can compensate for the phase detuning and achieve high growth rate, large gain bandwidth, and efficient, sustained energy transfer even in steep density gradient. A theoretical model based upon the three-wave coupling equations --- incorporating both the chirp effect and density detuning --- is constructed for the first time. An effective growth rate is analytically derived, demonstrating that the pulse can still be amplified even when it deviates from ideal resonance conditions. Numerical results obtained from three-wave coupling equations and PIC simulations have verified the effectiveness of both the theoretical model and the cFRA scheme, demonstrating its capability of generating multi-PW ultra-short pulses at different wavelengths. A parameter window for the chirp rate and density scale length suitable for robust cFRA performance is presented, which may be applied for experimental design. 
Notably, the cFRA scheme eliminates the requirement for plasma density uniformity in Raman amplification while simultaneously achieving higher output than that found in uniform plasma. It not only inherits the advantages of the Raman schemes with high growth rate, high tunability of seed wavelengths, and high robustness of forward scattering against kinetic and temperature effects, but also acquires immunity to nonuniformity similar to that of the Brillouin schemes. This addresses the long-standing limitations that have hindered Raman schemes over the past two or three decades, thereby advancing plasma-based light amplification toward practical applications.

This work is supported by the National Natural Science Foundation of China (Grant Nos. 125B2116, 12135009 and 12595362). Numerical simulations were performed on Computer $\pi$2.0 in the Center for High Performance Computing at Shanghai Jiao Tong University.


\bibliography{apssamp}

\subsection*{End Matter}
\begin{figure}[t]
\centering
\includegraphics[width=0.3\textwidth]{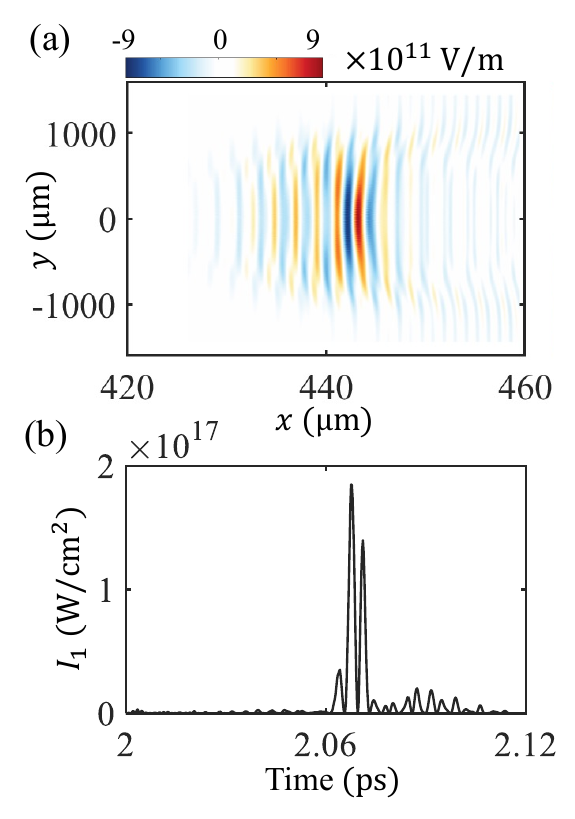}
\caption{2D PIC simulation results of cFRA scheme. (a) The spatial distribution of the electric fields of the amplified seed pulse. (b) The final output seed pulse intensity as a function of time.}
\label{2d}
\end{figure}

In 2D PIC simulations of the cFRA scheme, the ability to generate PW-class pulses is demonstrated, with a seed pulse initially at $80\,\mathrm{GW}$ amplified to $4.0\,\mathrm{PW}$. In this scenario, a chirped seed pulse of $1.6\,\mathrm{\mu m}$ central wavelength is adopted, featuring an initial intensity $10^{12}\,\mathrm{W/cm^2}$, duration $133\,\mathrm{fs}$, chirp rate $1.3\times10^{27}\,\mathrm{rad/s^2}$, and transverse spot radius $1.6\,\mathrm{mm}$ with the Gaussian transverse profile. The corresponding bandwidth is $14\%$. The pump pulse wavelength is $1.0\,\mathrm{\mu m}$, with initial intensity $3\times 10^{16}\,\mathrm{W/cm^2}$, duration $233\,\mathrm{fs}$ and the same spot size as the seed. The background plasma features a linear density ramp, with $x_0=250\,\mathrm{\mu m}$, and $L=450\,\mathrm{\mu m}$. As shown in Fig.~\ref{2d}, the final output intensity of the seed pulse increases to $1.8\times 10^{17}\,\mathrm{W/cm^2}$ and compressed to $8\,\mathrm{fs}$ or in $1.5$ cycles. The radius of the output seed is reduced to $900\,\mathrm{\mu m}$, due to the fact that the seed intensity in the outer region is too small to be amplified effectively, which is common and has been observed in previous backward Raman amplification experiments.  Meanwhile, no significant splitting and filaments are observed in both the longitudinal and transverse distributions of the seed pulse during the amplification, indicating that the cFRA has a higher tolerance to various instabilities. Compared to the FRA scheme which may even require a two-stage plasma density distribution to reduce the potential instabilities, cFRA reaches higher output intensity, and exhibits a higher contrast ratio with a suppressed pre-pulse. 

\begin{figure}[H]
\centering
\includegraphics[width=0.32\textwidth]{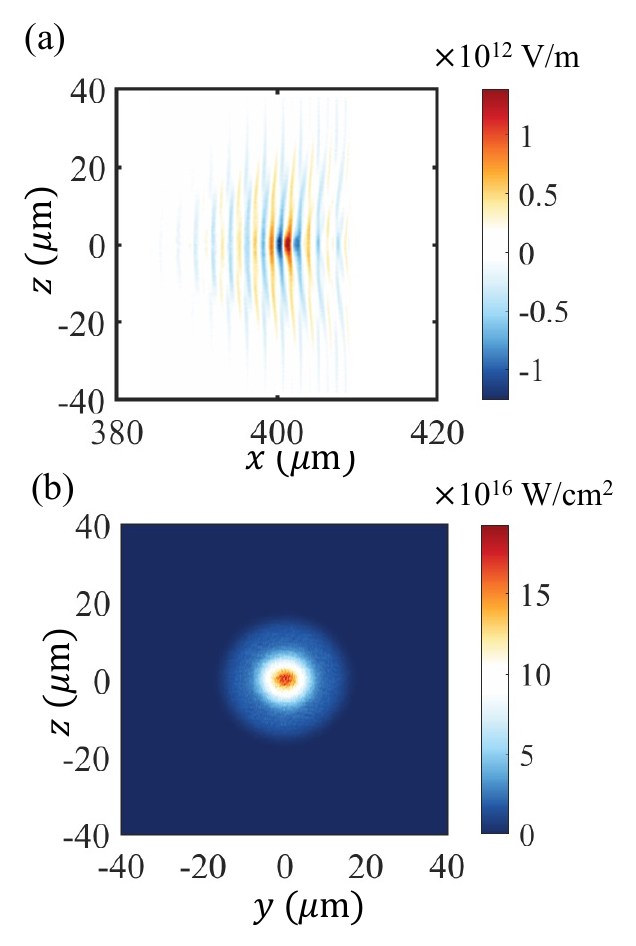}
\caption{3D PIC simulation results of cFRA scheme. (a) The spatial distribution of the amplified seed pulse electric field $E_y$ in the $x–z$ plane. (b) The spatial distribution of the amplified seed pulse intensity in the transverse plane.}
\label{3d}
\end{figure}

To further demonstrate the effectiveness of the cFRA scheme in a realistic geometry, a 3D PIC simulation is conducted with a smaller spot radius of $40\,\mathrm{\mu m}$, owing to the limitation by the large computational resources required in the 3D case. As shown in Fig.~\ref{3d}, the output chirped seed pulse is amplified beyond $10^{17}\,\mathrm{W/cm^2}$  in the inhomogeneous plasma effectively, and the duration is compressed to nearly single cycle without obvious development of the negative instabilities.

\end{document}